\documentclass[prd,twocolumn,eqsecnum,showpacs]{revtex4}
\usepackage{bm} 
\usepackage{amssymb}
\begin{document}
\title{Mass loss by a scalar charge in an expanding universe}     
\author{Lior M.~Burko}
\affiliation{Department of Physics, University of Utah, Salt Lake
City, Utah 84112}
\affiliation{Theoretical Astrophysics, California Institute of
Technology, Pasadena, California 91125} 
\author{Abraham I.~Harte}
\affiliation{Theoretical Astrophysics, California Institute of
Technology, Pasadena, California 91125} 
\author{Eric Poisson}
\affiliation{Department of Physics, University of Guelph, Guelph, 
  Ontario, Canada N1G 2W1}
\date{\today}  
\begin{abstract}
We study the phenomenon of mass loss by a scalar charge --- a point
particle that acts a source for a noninteracting scalar field --- in
an expanding universe. The charge is placed on comoving world lines 
of two cosmological spacetimes: a de Sitter universe, and a
spatially-flat, matter-dominated universe. In both cases, we find that
the particle's rest mass is not a constant, but that it changes in
response to the emission of monopole scalar radiation by the
particle. In de Sitter spacetime, the particle radiates all of its
mass within a finite proper time. In the matter-dominated cosmology, 
this happens only if the charge of the particle is sufficiently large;
for smaller charges the particle first loses some of its mass, but
then regains it all eventually. 
\end{abstract}
\pacs{04.25.-g, 95.30.Cq, 98.80.-k} 
\maketitle

\section{Introduction and summary} \label{intro}  

In this paper we study the phenomenon of mass loss by a scalar charge
in an expanding universe. By scalar charge we mean a pointlike 
particle carrying a charge $q$ that acts as a source for a massless scalar
field $\Phi$. The theory describing the particle-field system is a
close analogue to standard electromagnetic theory. It comes with a 
wave equation for the field, and equations of motion for the
particle; both equations are linear in $\Phi$. The equations predict
that the mass of the particle is not a constant, but that it changes 
dynamically as the particle moves in a curved spacetime \cite{Q}. This
comes about because the four-force produced by the scalar field, 
which is proportional to the gradient of $\Phi$, is not orthogonal to
the particle's four-velocity. Although such an unusual situation was
identified in the past \cite{thirring}, it has attracted surprisingly
little attention, and its consequences are well worth exploring.       

The usual notion of a particle's rest mass is that it is an invariant 
quantity that always stays constant. Further, one normally assumes
that all electrons, say, have the same mass, which is then a
fundamental constant of nature. Indeed, the whole concept of an
elementary particle tends to imply rest-mass universality, and
rest-mass conservation. It is also usually taken for granted that the
properties of elementary particles --- such as mass, charge, or spin
--- are independent of the cosmological parameters. While suggestions
were made to link cosmology to the properties of elementary particles 
(notably the large-numbers hypothesis \cite{ln1,ln2,ln3,ln4,ln5}),
this idea has not gained much popularity.    

We show here that such a link is unavoidable for scalar charges ---
a scalar particle at rest in an expanding universe possesses a
changing mass whose dynamics is directly coupled to the cosmology.  
We display this connection for two spacetimes: de Sitter, and a
spatially-flat matter-dominated cosmology. A de Sitter universe 
provides a reasonable description of the inflationary epoch of our own  
universe, and the flat matter-dominated universe adequately 
models our universe's present epoch (in the absence of dark energy).  
Thanks to the simplicity of these spacetimes, our mathematical
treatment of the mass-change phenomenon is exact: our calculations
involve no approximations. The phenomenon is not exclusive to the two
spacetimes considered here: A similar effect occurs in flat
($1+1$)-dimensional and ($2+1$)-dimensional Minkowski spacetimes, and
this is discussed in a separate paper \cite{burko-flat}. The effect,
however, does not occur in ($3+1$)-dimensional Minkowski spacetime,
nor in the Kerr family of spacetimes \cite{burko-liu}.  

For both cosmologies we calculate the changing mass of a scalar
charge taken to be comoving with the cosmological fluid. We
assume that the particle starts with a finite mass $m_0$ in the finite  
past, and we determine how the mass behaves thereafter; its behavior
depends on $m_0$ and $q$, as well as the cosmological
parameters. For
our matter-dominated universe, we find that two types of behaviors are 
possible, depending on the size of $q$ relative to a combination of
other parameters. In the first scenario ($q$ small), the particle
first loses a fraction of its mass, but it then regains it all
eventually. In the second scenario ($q$ large), the particle loses all
of its mass within a finite proper time. For de Sitter spacetime,
only the second scenario applies, and the particle loses mass at a
constant rate.     

The mechanism behind the mass-change phenomenon is easy to identify:
the scalar charge emits monopole waves, and the energy carried off
by the radiation is taken from the particle's rest mass. (As we show
in the paper, this interpretation is precise for de Sitter spacetime,
which admits a timelike Killing vector. It is not precise, but still
loosely correct, in the case of our matter-dominated universe, which
does not admit a timelike Killing vector.) The radiative process is 
fueled by the expansion of the universe (which provides the required 
dynamics), and its monopolar nature reflects the spherical symmetry of
the problem. This implies that the mass-change phenomenon should not
be expected for fields of higher spins: monopole waves could not 
be produced in such cases. Instead, the radiation would
necessarily be associated with higher multipole moments, and it could
not be produced by a changing rest mass. For example, the four-force 
produced by an electromagnetic field is always orthogonal to the 
four-velocity, and the rest mass of an electric charge is always
constant.  

The classical framework adopted here does not allow us to predict what
happens when a particle has radiated all of its mass. To avoid a
runaway regime of negative rest masses, we assume that the scalar
charge simply disappears when its mass drops to zero. This 
fix is imposed without a proper justification, but it seems
reasonable, and it does not appear to violate any known law of
physics. In this regard we remark that the theory adopted here to
describe the particle-field system does not automatically enforce
scalar charge conservation. We assume, for simplicity, that $q$ stays    
constant as long as the particle continues to exist, but that it jumps  
abruptly to zero when the particle has radiated all of its mass. It
would be interesting to consider more sophisticated models in which
the scalar charge, as well as the mass, would be dynamically
changing. We shall not, however, pursue this here. 

We note that our notion of a scalar charge --- a pointlike particle
acting as a source for a massless scalar field --- is different from
other models which also describe scalar particles. For example, Seidel 
and Suen \cite{suen-seidel-91} considered a soliton which is made of a
massive, sourceless scalar field. (For the Seidel-Suen soliton it is
important that the field be massive: first, the frequency of the
soliton-star oscillations vanishes without the mass term, and the
solution becomes unstable; second, from the studies of gravitational
critical phenomena we know that a massless scalar field either
disperses to infinity or collapses to a black hole
\cite{choptuik-93}.) The Seidel-Suen soliton is very
different from our notion of a scalar charge. It would be interesting, 
however, to investigate how the Seidel-Suen soliton behaves in a
cosmological spacetime. 

We begin in Sec.~\ref{action} with the presentation of an action
principle for the particle-field system, and a derivation of the
equations of motion. In Sec.~\ref{cos} we describe a class of
cosmological spacetimes that includes the two cosmologies of interest
to us. In Sec.~\ref{green} we calculate the retarded Green's function 
for a scalar field living in these two spacetimes. An alternative 
derivation, specific to de Sitter spacetime, is described in Appendix
A. In Sec.~\ref{field} we compute the field of a scalar charge at rest
in the two spacetimes, and show that the particle radiates monopole
waves. The mass loss is computed in Sec.~\ref{mass}. Finally,
Sec.~\ref{discussion} contains a discussion of energy conservation in
de Sitter spacetime, and an examination of the implications of the
mass-loss effect on the abundance of scalar charges in our own
universe. We also consider how our purely classical treatment might be
related to a fundamental quantum theory of radiating scalar
charges. Throughout the paper, except when stated otherwise, we use
geometrized units in which $c = G = 1$.     

\section{Action principle and equations of motion} \label{action}

A particle of bare mass $m_0$ and scalar charge $q$ moves on a world 
line $z^\alpha(\lambda)$ in a curved spacetime with metric 
$g_{\alpha\beta}$; the world line is monotonically parameterized by  
$\lambda$. The particle creates a scalar field $\Phi$, and the
dynamics of the particle-field system is described by the action
principle  
\begin{eqnarray}
S &=& \int \biggl\{ -\frac{1}{8\pi} g^{\alpha\beta} \Phi_{,\alpha}
\Phi_{,\beta} - \int (m_0 - q \Phi)  
\nonumber \\
& & \mbox{} \times
\sqrt{-g_{\alpha\beta} \dot{z}^\alpha \dot{z}^\beta}\,
\frac{\delta_4(x-z(\lambda))}{\sqrt{-g}}\, d\lambda  
\biggr\} \sqrt{-g}\, d^4 x, 
\nonumber \\
\label{2.1}
\end{eqnarray}
where $\Phi_{,\alpha} \equiv \partial \Phi/\partial x^\alpha$, 
$\dot{z}^\alpha \equiv d z^\alpha/d\lambda$, $g$ is the metric
determinant, and $\delta_4(x-z)$ is a four-dimensional Dirac
distribution which satisfies $\int \delta_4(x-z) d^4 x = 1$ if
$z^\alpha$ is within the domain of integration. We note that the
action $S$ is invariant under a reparameterization of the world
line. After variation, it is convenient to set $\lambda$ equal to the
particle's proper time $\tau$, which is obtained by integrating $d\tau
= \sqrt{-g_{\alpha\beta} \dot{z}^\alpha \dot{z}^\beta}\, d\lambda$. We 
also note that the metric $g_{\alpha\beta}$ does not participate in
the dynamics; it is a prescribed tensor field in spacetime.     

Variation of the action with respect to $\Phi$ produces a linear wave
equation for the field,  
\begin{equation}
g^{\alpha\beta} \nabla_\alpha \nabla_\beta \Phi = -4\pi \mu,
\label{2.2}
\end{equation}
where  
\begin{equation} 
\mu(x) = q \int \frac{\delta_4(x-z(\tau))}{\sqrt{-g}}\, d\tau
\label{2.3}
\end{equation}
is the scalar charge density. Variation of the action with respect to
$z^\alpha(\lambda)$ produces equations of the motion for the particle, 
\begin{equation}
m(\tau) \frac{D u^\alpha}{d\tau} = q \bigl( g^{\alpha\beta} 
+ u^\alpha u^\beta \bigr) \Phi_{,\beta},
\label{2.4}
\end{equation}
where $u^\alpha = dz^\alpha/d\tau$ is the four-velocity, 
$D u^\alpha/d\tau \equiv du^\alpha/d\tau  
+ \Gamma^\alpha_{\ \beta\gamma} u^\beta u^\gamma$ the covariant 
acceleration, and 
\begin{equation}
m(\tau) = m_0 - q \Phi
\label{2.5}
\end{equation}
the dynamical mass of the particle. This last equation is
equivalent to the differential statement 
\begin{equation}
\frac{d m}{d\tau} = -q \Phi_{,\alpha} u^\alpha,
\label{2.6}
\end{equation}
which is independent of the bare mass $m_0$.   

We note that Eqs.~(\ref{2.2})--(\ref{2.4}) and (\ref{2.6}) also follow 
from an action principle proposed by Quinn \cite{Q}: 
\begin{eqnarray}
S' &=& \int \biggl\{ -\frac{1}{8\pi} g^{\alpha\beta} \Phi_{,\alpha}
\Phi_{,\beta} 
+ \int \biggl[ \frac{1}{2} m(\tau)\, g_{\alpha\beta} u^\alpha u^\beta  
\nonumber \\
& & \mbox{}
+ q \Phi \biggr]\, \frac{\delta_4(x-z(\tau))}{\sqrt{-g}}\, d\tau 
\biggr\} \sqrt{-g}\, d^4 x. 
\label{2.7}
\end{eqnarray}
This action involves only the dynamical mass $m(\tau)$, but it is not
invariant under a reparameterization of the world line.   

The actions $S$ and $S'$ both produce a dynamically changing mass for
a scalar charge. It is possible to construct an action that produces
a constant mass. For example, replacing $(m_0 - q\Phi)$ in
Eq.~(\ref{2.1}) by $m_0 \exp(-q\Phi/m_0)$ gives rise to the equations
of motion $m_0\, Du^\alpha/d\tau = q(g^{\alpha\beta} + u^\alpha
u^\beta) \Phi_{,\beta}$. The price to pay, however, is high, as
Eq.~(\ref{2.2}) must now be replaced by the nonlinear wave equation
$g^{\alpha\beta} \nabla_\alpha \nabla_\beta \Phi =
-4\pi \mu \exp(-q\Phi/m_0)$.      

As they stand, Eqs.~(\ref{2.4})--(\ref{2.6}) have only formal 
validity because the field $\Phi$ derived from Eqs.~(\ref{2.2})
and (\ref{2.3}) is singular on the world line. Quinn \cite{Q} has
shown that the singular part of the field does not affect the motion
of the particle, which is then governed entirely by the smooth (or
tail) part of the field. Quinn has thus produced regularized versions
of the particle's equations of motion: 
\begin{eqnarray}
m \frac{D u^\alpha}{d\tau} &=& \frac{1}{6}\, q^2 \bigl( 
R^\alpha_{\ \beta} u^\beta + u^\alpha R_{\beta\gamma} u^\beta u^\gamma
\bigr) \nonumber \\
& & \mbox{} 
+ q^2 \bigl( g^{\alpha\beta} + u^\alpha u^\beta \bigr) 
\int_{-\infty}^{\tau_-} G_{,\beta}(\tau,\tau')\, d\tau' 
\nonumber \\
& & \label{2.8}
\end{eqnarray}
and 
\begin{equation}
\frac{dm}{d\tau} = -\frac{1}{12}\, q^2 R - q^2 u^\alpha    
\int_{-\infty}^{\tau_-} G_{,\alpha}(\tau,\tau')\, d\tau'. 
\label{2.9}
\end{equation} 
Here, $R_{\alpha\beta}$ is the spacetime's Ricci tensor, and $R$ is
the Ricci scalar. The quantity $G(x,x')$ appearing inside the
integrals is the retarded Green's function associated with the scalar
wave equation (\ref{2.2}); it satisfies 
\begin{equation}
g^{\alpha\beta} \nabla_\alpha \nabla_\beta G(x,x') 
= -4\pi \frac{\delta_4(x-x')}{\sqrt{-g}},     
\label{2.10}
\end{equation}
where $x$ is identified with $z(\tau)$, the current position of the
particle, while $x'$ is identified with $z(\tau')$, the particle's
past position. The integrals extend over the entire past world line of 
the particle, from $\tau' = -\infty$ to (almost) the current time,
$\tau' = \tau^- \equiv \tau - \epsilon$, where $\epsilon$ is 
infinitesimally positive \cite{Q}. The integration is cut short to
avoid the singular behavior of the Green's function as $x'$ approaches
$x$; it involves only the smooth part of the Green's function, which
is often referred to as its ``tail part''. In Eqs.~(\ref{2.8}) and
(\ref{2.9}), the four-velocity $u^\alpha$ is evaluated at the current
time $\tau$, and the Green's function is differentiated with respect
to $x$ before making the identification $x = z(\tau)$.

We will evaluate and solve Eq.~(\ref{2.9}) for the dynamical mass of a
scalar charge at rest in an expanding universe --- in this case the
right-hand side of Eq.~(\ref{2.8}) vanishes and the particle follows a
geodesic of the spacetime. We will see that solving Eq.~(\ref{2.9})
produces a regularized version of Eq.~(\ref{2.5}).  

\section{Cosmological spacetimes}\label{cos}

For simplicity we consider spatially-flat cosmologies, and write the 
metric as 
\begin{equation}
ds^2 = a^2(\eta)(-d\eta^2 + dx^2 + dy^2 + dz^2), 
\label{3.1}
\end{equation}
in terms of a conformal time $\eta$. The spacetime is filled with a 
homogeneous fluid of density $\rho$, pressure $p$, and four-velocity 
\begin{equation}
u^\alpha = \frac{dx^\alpha}{d\tau} = (a^{-1},0,0,0), 
\label{3.2}
\end{equation}
where $\tau$ is proper time for comoving observers, related to the
conformal time by $d\tau = a(\eta)\, d\eta$. The behavior of 
the scale factor $a(\eta)$ is governed by an energy-conservation
equation, $(a^3 \rho)' + p(a^3)' = 0$, in which a prime indicates
differentiation with respect to $\eta$, and Raychaudhuri's equation, 
$a^{\prime 2} = (8\pi/3) \rho a^4$.

We shall consider power-law cosmologies, for which the scale factor
takes the simple form 
\begin{equation}
a(\eta) = C \eta^\alpha, 
\label{3.3}
\end{equation}
where $C$ and $\alpha$ are constants. For such cosmologies, the
density is given by 
\begin{equation}
\rho = \frac{3 \alpha^2}{8\pi C^2}\, \frac{1}{\eta^{2\alpha+2}}, 
\label{3.4}
\end{equation}
the pressure by 
\begin{equation}
p = \frac{2-\alpha}{3\alpha}\, \rho, 
\label{3.5}
\end{equation}
and the Ricci scalar by 
\begin{equation}
R = \frac{6 \alpha(\alpha - 1)}{C^2}\, \frac{1}{\eta^{2\alpha + 2}}. 
\label{3.6}
\end{equation} 

Two special cases, which constitute an equivalence class in a sense to 
be described below, will be of interest to us. The first is a
matter-dominated cosmology characterized by $\alpha = 2$, which
produces a vanishing pressure. The second is a de Sitter cosmology
characterized by $\alpha = -1$, which produces a constant density and 
a pressure $p = -\rho$. In Table I we summarize the properties of
these cosmologies.  

\begin{table} 
\caption{The cosmological models considered in this paper. In both
  cases the scale factor is given by $a(\eta) = C \eta^\alpha$, and
  proper time is defined by $d\tau = a(\eta)\, d\eta$. For both
  cosmologies, the table displays the parameter $\alpha$, the scale
  factor (in terms of both $\eta$ and $\tau$), the relations
  $\tau(\eta)$ and $\eta(\tau)$, the density $\rho(\eta)$, the
  pressure-to-density ratio, and the Ricci scalar $R(\eta)$. In the
  spatially-flat, matter-dominated cosmology, both $C$ and $\eta$ are
  positive, and the universe is expanding. In the de Sitter cosmology,
  both $C$ and $\eta$ are negative, and $a(\eta) = |C|/(-\eta)$ also
  describes an expanding universe. (The maximally extended de Sitter
  spacetime also includes a preceding contracting phase which we do
  not consider here.)}  
\begin{ruledtabular}
\begin{tabular}{lll}
cosmology    & matter dominated       & de Sitter             \\
\hline                                                     
$\alpha$     & $2$                    & $-1$                  \\
$a(\eta)$    & $C \eta^2$             & $|C|/(-\eta)$         \\
$a(\tau)$    & $C(3\tau/C)^{2/3}$     & $|C| e^{\tau/|C|}$    \\
$\tau(\eta)$ & $C\eta^3/3$            & $-|C| \ln(-\eta)$     \\
$\eta(\tau)$ & $(3\tau/C)^{1/3}$      & $-e^{-\tau/|C|}$      \\
$\rho$       & $3/(2\pi C^2 \eta^6)$  & $3/(8\pi C^2)$        \\
$p/\rho$     & $0$                    & $-1$                  \\
$R$          & $12/(C^2 \eta^6)$      & $12/C^2$
\end{tabular} 
\end{ruledtabular}
\end{table} 

The scalar wave equation (\ref{2.2}) can be simplified if we introduce 
the auxiliary field variable $\psi$ defined by 
\begin{equation}
\Phi(\eta,\bm{x}) = \frac{1}{a(\eta)}\, \psi(\eta,\bm{x}), 
\label{3.7}
\end{equation}
in which $\bm{x} = (x,y,z)$ represents the spatial coordinates. This
new variable satisfies  
\begin{equation}
\biggl[ - \frac{\partial^2}{\partial \eta^2} + \nabla^2 +
\frac{\alpha(\alpha-1)}{\eta^2} \biggr] \psi = -4\pi a^3 \mu, 
\label{3.8}
\end{equation} 
where $\nabla^2 = \partial^2/\partial x^2 + \partial^2/\partial y^2 
+ \partial^2/\partial z^2$ is the flat-space Laplacian operator. On
the basis of Eq.~(\ref{3.8}) we can state the following 
interesting property: Two differing cosmological models with identical
values for $\beta \equiv \frac{1}{2} \alpha(\alpha-1)$ will produce 
identical reduced fields $\psi$, provided that the source term $a^3
\mu$ is the same in both cases, and that $\psi$ starts with the same
initial conditions. In this specific sense, we may say that two 
cosmological models with equal $\beta$ are ``equivalent''. For a
prescribed value of $\beta$, the two equivalent cosmological models
will be characterized by $\alpha = \alpha_\pm$, where
\begin{equation}
\alpha_\pm = \frac{1}{2} \pm \frac{1}{2} \sqrt{1+8\beta}. 
\label{3.9}
\end{equation}
Notice in particular that if $\beta = 1$, then $\alpha_+ = 2$ and  
$\alpha_- = -1$, and we obtain the cosmological models summarized in
Table I. These models are therefore equivalent in the sense adopted
here.      

\section{Green's function} \label{green}

In order to find solutions to Eq.~(\ref{2.10}), we factorize the Green's
function according to 
\begin{equation}
G(x,x') = \frac{1}{a(\eta)a(\eta')}\, g(x,x'),  
\label{4.1}
\end{equation}
which produces a reduced version of Green's equation, 
\begin{equation}
\biggl( - \frac{\partial^2}{\partial \eta^2} + \nabla^2 +
\frac{2\beta}{\eta^2} \biggr) g(x,x') = 
-4\pi \delta(\eta - \eta') \delta_3(\bm{x} - \bm{x'})\, .   
\label{4.2}
\end{equation} 
In the following two subsections we shall solve Eq.~(\ref{4.2}) using 
two different methods: the first relies on Hadamard's theory, 
and the second is based on a mode decomposition of Green's
equation. An alternative derivation, specific to de Sitter spacetime,
is described in Appendix A. 

\subsection{Solution by Hadamard ansatz}  

For $\beta = 0$ (flat spacetime), the retarded solution to
Eq.~(\ref{4.2}) is 
\begin{equation}
g^{\rm flat}(x,x') = \frac{\delta(u)}{|\bm{x}-\bm{x'}|},
\label{4.3}
\end{equation}
where $u = \eta - \eta' - |\bm{x}-\bm{x'}|$ is retarded
time; we see that the flat-spacetime Green's function has support only
on the past light cone of the field point $x$. Relying on Hadamard's
general theory \cite{H,deWB}, we expect that $g(x,x')$ will have
support inside the light cone as well, and we write it as 
\begin{equation}
g(x,x') = g^{\rm flat}(x,x') + B(x,x') \theta(u), 
\label{4.4}
\end{equation}
where $\theta(u)$ is the Heaviside step function and $B(x,x')$ a
two-point function to be determined, but which is known to be smooth 
when $u=0$. [Because the right-hand side of Eq.~(\ref{4.2}) is nothing
more than a flat-spacetime $\delta$-function, there is no need to
modify the $\delta(u)$ part of the Green's function: it is simply
equal to the flat-spacetime result. This assumption is justified by  
the following.] Substituting Eq.~(\ref{4.4}) into Eq.~(\ref{4.2}), 
we find that $g^{\rm flat}$ takes care of the four-dimensional 
$\delta$-function, and that the remainder vanishes as a distribution 
if $B$ satisfies 
\begin{equation}
(\square + V) B = 0 
\label{4.5}
\end{equation}
and 
\begin{equation}
B_{,\alpha} (x - x')^\alpha + B - \frac{1}{2}\, V = {\cal O},  
\label{4.6}
\end{equation}
where $\square = -\partial^2/\partial \eta^2 + \nabla^2$ is the
flat-spacetime d'Alembertian operator and $V(x) \equiv 2
\beta/\eta^2$. The right-hand side of Eq.~(\ref{4.6}) is a priori
arbitrary, but it must vanish when $u=0$; it is also constrained by
the fact that a simultaneous solution to both Eqs.~(\ref{4.5})
and (\ref{4.6}) must exist.  As a consequence of Eq.~(\ref{4.6}) and
the fact that $B$ is smooth at $u=0$, we establish the coincidence
limit   
\begin{equation}
\lim_{x \to x'} B(x,x') = \frac{1}{2}\, V. 
\label{4.7}
\end{equation}       
This boundary condition allows us to solve Eqs.~(\ref{4.5}) and
(\ref{4.6}) uniquely.  

For simplicity we shall set the right-hand side of Eq.~(\ref{4.6}) to 
zero. As we shall see, this will produce the condition $\beta = 1$,
and we shall therefore be restricted to the cosmological models
summarized in Table I. 

Let $x$ and $x'$ be fixed points in spacetime, and let the 
relations $x^{\prime\prime\alpha} = x^{\prime \alpha} + (\eta''-\eta')  
n^\alpha$ describe a straight line going from $x'$ to $x$ as the
parameter $\eta''$ ranges from $\eta'$ to $\eta$; the tangent vector
$n^\alpha = dx^\alpha/d\eta''$ is normalized so that $n^\eta =
1$. Rewrite Eq.~(\ref{4.6}) in terms of the variable $x''$ and note
that the derivative of $B$ in the direction of $(x''-x')^\alpha$ is
equal to $\eta'' dB/d\eta''$. Equation (\ref{4.6}) can therefore be 
re-expressed as   
\begin{equation}
\eta'' \frac{d B}{d\eta''} + B = \frac{1}{2} V(\eta''),      
\label{4.8}
\end{equation}
and this can be straightforwardly integrated. The solution that
satisfies Eq.~(\ref{4.7}) is  
\begin{equation}
B(x,x') = \frac{1}{2(\eta - \eta')} \int_{\eta'}^\eta V(\eta'')\,
d\eta''. 
\label{4.9} 
\end{equation} 
With $V = 2\beta/\eta^2$ we have that $B(x,x') = \beta/(\eta \eta')$.  

Thus far we have generated a solution to Eq.~(\ref{4.6}) only. (Recall
that we have set ${\cal O} = 0$ in this equation.) We must now
check that this is also a solution to Eq.~(\ref{4.5}). Because both
$V$ and $B$ are proportional to $\beta$, it is easy to see that this
produces a constraint on the value of $\beta$. As we have indicated
previously, $B(x,x') = \beta/(\eta\eta')$ is a solution to both
Eqs.~(\ref{4.5}) and (\ref{4.6}) if and only if $\beta = 1$.   

Our conclusion is that for $\beta = 1$ (which implies either 
$\alpha=2$ or $\alpha = -1$), the retarded solution to 
Eq.~(\ref{4.2}) is   
\begin{equation} 
g(x,x') =
\frac{\delta(\eta-\eta'-|\bm{x}-\bm{x'}|)}{|\bm{x}-\bm{x'}|} 
+ \frac{\theta(\eta-\eta'-|\bm{x}-\bm{x'}|)}{\eta \eta'}. 
\label{4.10} 
\end{equation} 
The cosmological models summarized in Table I are equivalent in the 
sense used before, but also in the sense that they come with the 
same reduced Green's function $g(x,x')$. Because the scale factors are
different, however, the actual retarded Green's function $G(x,x')$, 
given by Eq.~(\ref{4.1}), takes a distinct form in each spacetime.

\subsection{Solution by mode-sum} 

The method used in the preceding subsection to generate the retarded 
Green's function was limited to the case $\beta = 1$. Here we describe
an alternative method which could, if desired, be applied to any value
of $\beta$. Here we set $\beta = \frac{1}{2} l (l+1)$, and we notice
that this covers cosmological models characterized by either $\alpha = 
l+1$ or $\alpha = -l$. We will consider the case $l = 1$ in detail,
and reproduce Eq.~(\ref{4.10}).     

We expand the reduced Green's function $g(x,x')$ in terms of
plane-wave solutions to Laplace's equation, 
\begin{equation}
g(x,x') = \frac{1}{(2\pi)^3}\, \int \tilde{g}(\eta,\eta';\bm{k})\,
e^{i \bm{k} \cdot (\bm{x} - \bm{x'})}\, d^3 k,
\label{4.11}
\end{equation}
and we substitute this into Eq.~(\ref{4.2}). The result, after also
Fourier transforming $\delta_3(\bm{x}-\bm{x'})$, is an ordinary
differential equation for $\tilde{g}(\eta,\eta';\bm{k})$: 
\begin{equation}
\biggl[ \frac{d^2}{d\eta^2} + k^2 - \frac{l(l+1)}{\eta^2} \biggr]
\tilde{g} = 4\pi \delta(\eta - \eta'), 
\label{4.12}
\end{equation}
where $k^2 = \bm{k} \cdot \bm{k}$. To generate the retarded
Green's function we set 
\begin{equation}
\tilde{g}(\eta,\eta';\bm{k}) = \theta(\eta-\eta')\,
\hat{g}(\eta,\eta';k),  
\label{4.13}
\end{equation}
in which we indicate that $\hat{g}$ depends only on the modulus of the
vector $\bm{k}$. Substitution of Eq.~(\ref{4.13}) into
Eq.~(\ref{4.12}) reveals that $\hat{g}$ must satisfy the homogeneous 
version of Eq.~(\ref{4.12}),
\begin{equation}
\biggl[ \frac{d^2}{d\eta^2} + k^2 - \frac{l(l+1)}{\eta^2} \biggr]
\hat{g} = 0, 
\label{4.14}
\end{equation}
together with the boundary conditions 
\begin{equation}
\hat{g}(\eta=\eta';k) = 0, \qquad
\frac{d\hat{g}}{d\eta}(\eta=\eta';k) = 4\pi. 
\label{4.15}
\end{equation} 
Substitution of Eq.~(\ref{4.13}) into Eq.~(\ref{4.11}) and integration
over the angular variables associated with $\bm{k}$ yields 
\begin{equation}
g(x,x') = \frac{\theta(\eta-\eta')}{2\pi^2 R} \int_0^\infty 
\hat{g}(\eta,\eta';k)\, k \sin(kR)\, dk, 
\label{4.16}
\end{equation}
where $R \equiv |\bm{x}-\bm{x'}|$. 

For $l = 0$, the unique solution to Eqs.~(\ref{4.14}) and (\ref{4.15})
is 
\begin{equation}
\hat{g}_{l=0}(\eta,\eta';k) = \frac{4\pi}{k} \sin (k \Delta \eta), 
\label{4.17}
\end{equation}
where $\Delta \eta = \eta - \eta'$. Substituting this into
Eq.~(\ref{4.16}) returns the flat-spacetime Green's function of
Eq.~(\ref{4.3}). To derive this we make use of the identity 
\begin{equation}
\frac{2}{\pi} \int_0^\infty \sin(\omega x) \sin(\omega x')\, d\omega
= \delta(x-x') - \delta(x+x'), 
\label{4.18}
\end{equation}
and we note that the second $\delta$-function is eliminated by the
step function $\theta(\Delta \eta)$ in Eq.~(\ref{4.16}).   

If $l$ is an integer different from zero, Eq.~(\ref{4.14}) can be
solved in terms of spherical Bessel functions. (If $l$ is not an
integer, the equation can still be solved in terms of ordinary Bessel
functions.) It is simpler, however, to generate solutions with the
ladder operator $L_l \equiv -d/d\eta + (l+1)/\eta$. This works as
follows: Suppose that we already have $\hat{g}_l$, a solution to
Eq.~(\ref{4.14}) with a given value of $l$; then $\hat{g}_{l+1} = 
L_l \hat{g}_l$ is a solution to Eq.~(\ref{4.14}) with $l$ replaced by 
$l+1$. By repeated application of the ladder operator, a solution to 
Eq.~(\ref{4.14}) with any integer value of $l$ can be obtained from a
seed solution $\hat{g}_0$.  

We use this procedure to generate a solution $\hat{g}_{l=1}$ that
satisfies the boundary conditions (\ref{4.15}). For seed solutions
we use the set $\{ \sin (k \Delta \eta), \cos (k \Delta
\eta)\}$. After application of $L_0$, we find that $\hat{g}_{l=1}$
must be given by a superposition of the linearly independent solutions 
$\cos (k \Delta \eta) - (k \eta)^{-1} \sin (k \Delta \eta)$ and 
$\sin (k \Delta \eta) + (k \eta)^{-1} \cos (k \Delta \eta)$. The
coefficients are arbitrary functions of $\eta'$, and after imposing 
Eqs.~(\ref{4.15}), we find that the appropriate combination is      
\begin{eqnarray}
\hat{g}_{l=1}(\eta,\eta';k) &=& \frac{4\pi}{k}\, \biggl[ 
\biggl( 1 + \frac{1}{k^2 \eta \eta'} \biggr) \sin(k \Delta \eta) 
\nonumber \\ 
& & \mbox{} 
- \frac{\Delta \eta}{k \eta \eta'}\, \cos(k \Delta \eta) 
\biggr]. 
\label{4.19}
\end{eqnarray} 
Substituting this into Eq.~(\ref{4.16}) and using Eq.~(\ref{4.18})
yields 
\begin{equation}
g_{l=1}(x,x') = \frac{\delta(\Delta \eta - R)}{R} 
+ \frac{\theta(\Delta \eta)}{\eta \eta'}\, I(\Delta \eta,R), 
\label{4.20}
\end{equation}
where 
\begin{eqnarray}
I(\Delta \eta,R) &=& \frac{2}{\pi R} \int_0^\infty
\frac{\sin(kR)}{k^2}\, \bigl[ \sin(k\Delta\eta) 
\nonumber \\
& & \mbox{} 
- (k\Delta \eta) \cos(k \Delta \eta) \bigr]\, dk.  
\label{4.21}
\end{eqnarray} 
We evaluate this by integrating the first term by parts; this gives 
\begin{eqnarray} 
I(\Delta \eta,R) &=& \frac{2}{\pi} \int_0^\infty \frac{1}{k}\,  
\sin(k \Delta \eta) \cos(k R)\, dk 
\nonumber \\
&=& \theta(\Delta \eta - R).  
\label{4.22}
\end{eqnarray}
We therefore arrive at 
\begin{equation}
g_{l=1}(x,x') = \frac{\delta(\Delta \eta - R)}{R} 
+ \frac{\theta(\Delta \eta - R)}{\eta \eta'}, 
\label{4.23}
\end{equation}
which agrees with Eq.~(\ref{4.10}). 

The method described here could be extended to other values of $l$,
but we shall not pursue this here.  

\section{Field of a stationary charge} \label{field}

A particular solution to Eq.~(\ref{2.2}) is
\begin{equation}
\Phi(x) = \int G(x,x') \mu(x') \sqrt{-g'}\, d^4 x', 
\label{5.1}
\end{equation}
where $G(x,x')$ is the retarded Green's function of Eq.~(\ref{2.10})
and $g'$ is the determinant of the metric evaluated at $x'$; the
integration is over the entire spacetime manifold. In terms of the
reduced variables $\psi(x)$ and $g(x,x')$ --- cf.~Eqs.~(\ref{3.7}) and
(\ref{4.1}) --- we have
\begin{equation}
\psi(x) = \int g(x,x') a^3(\eta') \mu(x')\, d^4 x'. 
\label{5.2}
\end{equation} 
In this section we evaluate this for a stationary scalar charge. 

The scalar charge density of a point particle is given by
Eq.~(\ref{2.3}). For a particle at rest (comoving) in an expanding
universe with metric (\ref{3.1}), the four-velocity is given by
Eq.~(\ref{3.2}), and Eq.~(\ref{2.3}) gives $a^3(\eta)
\mu(\eta,\bm{x}) = q\, \delta_3(\bm{x})$, if we choose $\bm{x} = 
\bm{0}$ to represent the particle's position. (Because the spacetime
is homogeneous, there is no loss of generality in this choice.) 
Substituting this into Eq.~(\ref{5.2}) gives  
\begin{equation}
\psi(\eta,\bm{x}) = \int q\, g(\eta;\bm{x},\eta';\bm{0})\,
d\eta', 
\label{5.3}
\end{equation}
where the reduced Green's function is given by Eq.~(\ref{4.10}). 

For a matter-dominated cosmology ($\alpha=2,\beta=1$ --- see Table I),
the integration starts at $\eta' = 0$ where the Green's function is 
singular, and the integral of Eq.~(\ref{5.3}) is logarithmically
divergent. For a de Sitter cosmology ($\alpha=-1,\beta=1$), the
integration starts at $\eta' = -\infty$ and the integral is also 
logarithmically divergent. To avoid this pathology, we assume that the 
scalar charge came into being in the finite past, and we let 
\begin{equation}
q \to q\, \theta(\eta - \eta_0)
\label{5.4}
\end{equation}
in Eq.~(\ref{5.3}), which becomes 
\begin{equation}
\psi(\eta,\bm{x}) = q \int_{\eta_0}^\eta
g(\eta;\bm{x},\eta';\bm{0})\, d\eta'. 
\label{5.5}
\end{equation}
The replacement of Eq.~(\ref{5.4}) describes the sudden creation of a
scalar charge at a time $\eta = \eta_0 \neq \{0,-\infty\}$, and this 
constitutes a simple cure for the pathology of Eq.~(\ref{5.3}). It is
also possible to let the charge adiabatically ``switch on'', but this
would needlessly lead to more complicated expressions. We recall that 
there is no law of charge conservation in this theory: a scalar charge
can be spontaneously created provided that a sufficient amount energy
is made available.           

Integration of Eq.~(\ref{5.5}), with the reduced Green's function of
Eq.~(\ref{4.11}), is elementary, and we obtain 
\begin{equation}
\psi(\eta,\bm{x}) = \frac{q}{r}\, \theta(\eta - r - \eta_0) 
\biggl[ 1 + \frac{r}{\eta} \ln\biggl(\frac{\eta - r}{\eta_0} 
\biggr) \biggr],  
\label{5.6}
\end{equation}
where $r = |\bm{x}|$ is the coordinate distance from the
origin. Equation (\ref{5.6}) describes the reduced scalar field of a
stationary charge at $\bm{x} = \bm{0}$; the full scalar field is
given by $\Phi(\eta,\bm{x}) = \psi(\eta,\bm{x})/a(\eta)$. The step
function in Eq.~(\ref{5.6}) indicates that the moment of charge
creation is registered at a time $\eta = \eta_0 + r$ by an observer at
a distance $r$ from the charge: the information travels at the speed
of light. Equation (\ref{5.6}) implies that a stationary scalar charge
in an expanding universe radiates monopole waves: While the reduced
scalar field $\psi$ is stationary in the immediate vicinity of the
charge, the reorganization of the field lines caused by the underlying
spacetime curvature (which is dynamical) produces radiation. The
expansion of the universe also participates directly in the dynamics
of the scalar field; this is reflected by the presence of the scale
factor in the relation $\Phi = \psi/a$.   

The field of Eq.~(\ref{5.6}) is singular at $\bm{x} = \bm{0}$, 
where the particle is located. We define a {\it renormalized local
field\/} $\psi_{\rm ren}$ by first removing the singular part of  
$\psi$ and then taking the limit $r\to 0$. This gives  
\begin{equation}
\psi_{\rm ren}(\eta) = \frac{q}{\eta}\, \theta(\eta-\eta_0) 
\ln \bigl( \eta/\eta_0 \bigr). 
\label{5.7}
\end{equation}
We also define $\Phi_{\rm ren}(\eta) \equiv 
\psi_{\rm ren}(\eta)/a(\eta)$; the physical significance of this
quantity will be revealed in the next section. 

\section{Mass loss} \label{mass}

In this section we evaluate Eq.~(\ref{2.9}) for a stationary scalar
charge in the two cosmological spacetimes described in Table I. For
this situation, the particle's four-velocity is given by
Eq.~(\ref{3.2}), and the relevant (smooth) part of the Green's
function is obtained from Eqs.~(\ref{4.1}) and (\ref{4.10}): 
\begin{equation}
G_{\rm smooth}(x,x') = \frac{1}{C^2}\, 
\frac{1}{(\eta \eta')^{\alpha+1}}.  
\label{6.1}
\end{equation}
We recall that the scale factor is given by $a(\eta) = C \eta^\alpha$,
where $C$ and $\alpha$ are constants; Eq.~(\ref{6.1}) is valid if and
only if $\alpha$ is restricted to the values $2$ (matter-dominated
cosmology) and $-1$ (de Sitter cosmology). We recall also that for
these cosmologies, the Ricci scalar is given by Eq.~(\ref{3.6}). 

For a stationary particle, $u^\alpha G_{,\alpha}$ reduces to 
$a^{-1} \partial_\eta G$, and the proper-time integral of 
Eq.~(\ref{2.9}) can be expressed as an integral over $d\eta' =
a^{-1}(\eta')\, d\tau'$. This yields
\begin{eqnarray}
\frac{dm}{d\eta} &=& -\frac{1}{12} q^2 a(\eta) R(\eta) 
\nonumber \\
& & \mbox{}
- q^2 \int_{\eta_0}^\eta \partial_\eta G_{\rm smooth}(\eta,\eta')
a(\eta')\, d\eta' 
\label{6.2}
\end{eqnarray} 
for the rate of change of the particle's dynamical mass. Notice that
we have incorporated the cutoff of Eq.~(\ref{5.4}) into this
expression.  

\subsection{Matter-dominated cosmology} 

We now evaluate Eq.~(\ref{6.2}) for $\alpha = 2$. Substitution of
Eqs.~(\ref{6.1}) and (\ref{3.6}) yields  
\begin{equation}
\frac{dm}{d\eta} = -\frac{q^2}{C\eta^4}\, 
\bigl[ 1 - 3 \ln(\eta/\eta_0) \bigr],        
\label{6.3}
\end{equation}
and this can be immediately integrated:  
\begin{equation} 
m(\eta) = m_0 - \frac{q^2}{C\eta^3}\, \ln(\eta/\eta_0), 
\label{6.4}
\end{equation}
where $m_0 \equiv m(\eta_0)$. This result can be restated in terms of
the renormalized local field of Eq.~(\ref{5.7}): 
\begin{equation}
m(\eta) = m_0 - q \Phi_{\rm ren}(\eta), 
\label{6.5}
\end{equation}
which is analogous to Eq.~(\ref{2.5}).  

In order to analyze Eq.~(\ref{6.4}) we express it as 
\begin{equation}
\frac{m(\eta)}{m_0} = 1 - \frac{c}{x^3}\, \ln(x) 
\equiv f(x;c), 
\label{6.6}
\end{equation}
in terms of the rescaled quantities  
\begin{equation}
x = \frac{\eta}{\eta_0}, \qquad
c = \frac{q^2}{C m_0 {\eta_0}^3}. 
\label{6.7}
\end{equation}
The function $f(x;c)$ is defined in the interval $1 \leq x <
\infty$. It initially decreases from $f(1;c) = 1$ as $x$ increases,
and reaches its minimum value $f_{\rm min} = 1 - c/(3e)$ at $x =
e^{1/3} \simeq 1.3956$. Then $f(x;c)$ starts to increase, and it is
eventually restored to its original value, $f(x\to\infty;c) \to
1$. For $c \geq 3e \simeq 8.1548$, $f(x;c)<0$ in an interval around 
$x = e^{1/3}$.  

>From these considerations, we conclude that for $q^2 < 3 e C m_0
{\eta_0}^3$, the particle first loses mass, but eventually regains all
of it as $\eta \to \infty$. For $q^2 > 3 e C m_0 {\eta_0}^3$, on the
other hand, the particle radiates {\it all} of its mass in a time
shorter than $\Delta \eta = (e^{1/3}-1) \eta_0 \simeq 0.3956 \eta_0$;
as was conjectured in Sec.~\ref{intro}, this presumably signals the   
destruction of the scalar charge.    

\subsection{de Sitter cosmology} 

For $\alpha = -1$, $G_{\rm smooth}(\eta,\eta')$ is a constant, and the
integral term of Eq.~(\ref{6.2}) contributes nothing to the mass 
loss. Instead, this comes entirely from the Ricci-scalar term, and we
find 
\begin{equation}
\frac{dm}{d\tau} = -\frac{q^2}{C^2}.
\label{6.8}
\end{equation}
In terms of proper time $\tau$, mass is being lost at a constant 
rate. This result can also be expressed as 
\begin{equation}
m(\tau) = m_0 - \frac{q^2}{C^2}\, \bigl( \tau - \tau_0 \bigr), 
\label{6.9}
\end{equation}
where $m_0 \equiv m(\tau_0)$ is the initial mass of the scalar
charge, and the proper time $\tau_0$ is related to the conformal 
time $\eta_0$. (The relevant relations between proper and conformal
times are listed in Table I.) Here also we find that Eq.~(\ref{6.9})
can be simply restated in terms of the renormalized local field of 
Eq.~(\ref{5.7}):  
\begin{equation}
m(\tau) = m_0 - q \Phi_{\rm ren}(\tau);
\label{6.10}
\end{equation}
this again is analogous to Eq.~(\ref{2.5}). 

Our conclusion here is that the scalar charge will radiate {\it all}
of its mass, at a constant rate, within a proper time $\Delta \tau =
C^2 m_0/q^2$. This again signals the destruction of the scalar 
particle.  

\section{Discussion: Mass loss in de Sitter spacetime}
\label{discussion}  

\subsection{Energy Conservation}

The de Sitter spacetime possesses a maximal set of ten linearly
independent Killing vectors, and each one can be associated with a  
global conservation law. The spatial symmetries of the problem
immediately imply linear and angular momentum conservation, but the
explicit time dependence does not permit a hasty dismissal of  
energy conservation. Here we show that energy is globally conserved:
the energy radiated by a stationary scalar charge in de Sitter
spacetime is exactly equal to the mass lost. We note that such an
analysis cannot be adapted to the case of a matter-dominated
cosmology, because this spacetime does not admit a timelike Killing
vector.     

Let $T^{\alpha\beta}$ be the stress-energy tensor of the
particle-field system, and let $\xi^\alpha$ denote the timelike
Killing vector of de Sitter spacetime. By virtue of the equations
$T^{\alpha\beta}_{\ \ \ ;\beta} = 0$ and $\xi_{\alpha;\beta} +
\xi_{\beta;\alpha} = 0$ we find that the vector $j^\alpha =
-T^{\alpha}_{\ \beta} \xi^\beta$ is divergence free. By integrating
$j^\alpha_{\ ;\alpha} = 0$ over a bounded four-dimensional volume $V$
and using Gauss' theorem, we obtain the conservation statement 
\begin{equation}
E \equiv - \oint_{\partial V} T^{\alpha}_{\ \beta} \xi^\beta\,
d\Sigma_\alpha = 0, 
\label{6.11}
\end{equation}
where $\partial V$ is the volume's boundary, and $d\Sigma_\alpha$ is
an outward-directed surface element on $\partial V$; if
$(y^1,y^2,y^3)$ are coordinates intrinsic to $\partial V$, then
$d\Sigma_\mu = \varepsilon_{\mu\alpha\beta\gamma} 
(\partial x^\alpha/\partial y^1)(\partial x^\beta/\partial y^2)
(\partial x^\gamma/\partial y^3)\, d^3 y$, where  
$\varepsilon_{\mu\alpha\beta\gamma}$ is the totally antisymmetric  
Levi-Civita tensor.    

The total stress-energy tensor $T^{\alpha\beta}$ is obtained by
varying the action of Eq.~(\ref{2.1}) with respect to the metric. We 
express it as   
\begin{equation}
T^{\alpha\beta} = T^{\alpha\beta}_{\rm field} 
+ T^{\alpha\beta}_{\rm particle},
\label{6.12}
\end{equation}
where 
\begin{equation} 
T_{\alpha\beta}^{\rm field} = \frac{1}{4\pi} \biggl( 
\Phi_{,\alpha} \Phi_{,\beta} - \frac{1}{2}\, g_{\alpha\beta}\, 
\Phi^{,\mu} \Phi_{,\mu} \biggr) 
\label{6.13}
\end{equation} 
is the stress-energy tensor of the scalar field, and 
\begin{equation} 
T^{\alpha\beta}_{\rm particle} = \int m(\tau) u^\alpha(\tau)
u^\beta(\tau) \frac{\delta(x-z(\tau))}{\sqrt{-g}}\, d\tau 
\label{6.14}
\end{equation}
is the stress-energy tensor of the particle. 

We want to evaluate Eq.~(\ref{6.11}) when $\partial V$ consists of a  
cylindrical ``tube'' $B$ surrounding the particle's world line, closed off
by
``caps'' $C_1$ and $C_2$ (each of constant time) at both ends. It
would be inconvenient to
carry out this computation in the cosmological coordinates
$(\eta,x,y,z)$, because in these coordinates the metric is explicitly
time dependent, and the Killing symmetry is poorly represented. We
therefore prefer to use static coordinates $(t,r^*,\theta,\phi)$, in
which the metric is explicitly time independent. The transformation is   
\begin{eqnarray}
\eta &=& -\cosh(\kappa r^*) e^{-\kappa t}, \nonumber \\
& & \label{6.15} \\
(x,y,z) &=& \sinh(\kappa r^*) e^{-\kappa t} 
(\sin\theta \cos\phi, \sin\theta \sin\phi, \cos\theta), \nonumber  
\end{eqnarray} 
where $\kappa \equiv 1/|C|$, and in the new coordinates, the de
Sitter metric takes the form  
\begin{equation} 
ds^2 = \frac{-dt^2 + dr^{*2}}{\cosh^2(\kappa r^*)} + \kappa^{-2}
\tanh^2(\kappa r^*) \bigl( d\theta^2 + \sin^2\theta\, d\phi^2 \bigr). 
\label{6.16}
\end{equation} 
In these coordinates, $\xi^\alpha = \delta^\alpha_{\ t}$, and the Green's
function of Eqs.~(\ref{4.1}) and (\ref{4.10}) is  
\begin{equation} 
G(x,x') = \kappa \coth(\kappa r^*) \delta(t-t'-r^*) 
+ \kappa^2 \theta(t-t'-r^*)
\label{6.17}
\end{equation} 
if the source point is at the spatial origin of the coordinate
system. The field of a stationary charge at that position is given by   
\begin{equation} 
\Phi(t,r^*) = q \kappa \theta(t-r^*-t_0) \Bigl[ \coth(\kappa r^*) +
\kappa (t-r^*-t_0) \Bigr], 
\end{equation}
where $t_0$ denotes the time at which the scalar charge came into
being. Because $t$ is proper time for an observer at $r^* = 0$, the
dynamical mass of the scalar particle can be expressed as 
\begin{equation}
m(t) = m_0 - q^2 \kappa^2 (t-t_0), 
\label{6.18}
\end{equation}
where $m_0$ is the particle's mass at the time $t_0$; this follows
directly from Eq.~(\ref{6.9}).   

We choose the closed hypersurface $\partial V$ to be the union of a 
three-cylinder $B$ described by $(r^* = R, t_1 < t < t_2)$, a
spherical ball $C_1$ described by $(t = t_1, 0 < r^* < R)$, and
another ball $C_2$ described by $(t = t_2, 0 < r^* < R)$. After
decomposing the total stress-energy tensor as in Eq.~(\ref{6.12}), we
find that Eq.~(\ref{6.11}) becomes 
\begin{eqnarray} 
E &=& E_{\rm field}[B] + E_{\rm field}[C_2] - E_{\rm field}[C_1] 
\nonumber \\
& & \mbox{} + E_{\rm particle}[B] + E_{\rm particle}[C_2] - E_{\rm
particle}[C_1] 
\nonumber \\
&=& 0,  
\label{6.19}
\end{eqnarray}
where, for example, $E_{\rm field}[B] = - \int_{B} 
T^{\ \ \ \alpha}_{{\rm field} \beta} \xi^\beta\, d\Sigma_\alpha$. The 
minus signs in front of the terms associated with $C_1$ reflect the
fact that the future-directed surface element on $C_1$ points within
the region $V$, and therefore against the outward-directed surface
element on $\partial V$. 

Omitting all calculational details, we now present our results for the 
various quantities appearing in Eq.~(\ref{6.19}). First, we find that  
\begin{eqnarray}
E_{\rm field}[C_2] &=& E_{\rm field}[C_1] = \frac{1}{2} q^2 \kappa^2 
\nonumber \\ 
& & \mbox{} \times \int_0^R \bigl[ \tanh^2(\kappa r^*) 
+ \coth^2(\kappa r^*) \bigr]\, dr^*. 
\nonumber \\
& & \label{6.20}
\end{eqnarray}
This represents the field energy enclosed within a sphere of
coordinate radius $R$ centered at the origin. Because the field  
configuration is singular at $r^* = 0$, both 
$E_{\rm field}[C_2]$ and $E_{\rm field}[C_1]$ are formally
infinite. A proper derivation of global energy conservation should  
therefore involve a regularization procedure for the field's  
energy. Fortunately, equality of $E_{\rm field}[C_2]$ and 
$E_{\rm field}[C_1]$ ensures that independently of the details of the 
regularization procedure, these quantities cancel each other out on
the right-hand side of Eq.~(\ref{6.19}). (A technical requirement is
that the regularization procedure must not violate the
time-translational invariance of the field energy.) Second, we have  
\begin{equation}
E_{\rm field}[B] = q^2 \kappa^2 (t_2 - t_1), 
\label{6.21}
\end{equation}
which represents the energy radiated by the particle during the
time interval $t_1 < t < t_2$ (notice that this is independent of
$R$). Third,
\begin{equation}
E_{\rm particle}[C_2] = m(t_2), \qquad
E_{\rm particle}[C_1] = m(t_1),
\label{6.22}
\end{equation} 
and $E_{\rm particle}[B] = 0$: the energy carried by
the particle is equal to the current value of its mass.  

Substituting Eqs.~(\ref{6.21}) and (\ref{6.22}) into Eq.~(\ref{6.19}) 
yields 
\begin{equation}
E = q^2 \kappa^2 (t_2 - t_1) + m(t_2) - m(t_1)\, . 
\label{6.23}
\end{equation}
Using Eq.~(\ref{6.18}) reveals that energy is globally
conserved: the change in the mass is exactly equal to the energy
radiated, and $E = 0$.  

\subsection{Implications for this universe}

Can the mass-loss effect be used to explain the observed absence of 
scalar charges? We argue here that the answer is in the
affirmative. We will calculate an upper bound on the total 
number of scalar charges that might still exist today, and show that
it is very small, of the order of millions per galaxy. And we will
calculate a lower bound on the mass of a scalar charge, and show that
it is too large to permit the production of these particles in 
today's accelerators. We assume that scalar charges already existed at
the onset of the inflationary epoch, and that no additional scalar
charges were created since. 

The following discussion must be preceded by an important 
disclaimer. The mass-loss phenomenon was investigated in
Sec.~\ref{mass} within a framework in which the degrees of freedom
associated with the charge and its scalar field were both described by
classical physics. While we can hope that such a treatment may not be
blatantly at odds with a proper quantum description, the following
speculations regarding the fate of elementary scalar charges must be
regarded as more conjecture than definite prediction. In this regard we
can be encouraged by the fact that the classical treatment of
electromagnetic radiation reaction is often not a bad approximation to the
correct quantum description \cite{jackson}.   

A fundamental quantum theory of radiating scalar charges might be
based on the following considerations. The first ingredient would be a 
complex scalar field $\Phi_{\rm particle}$ whose quantum excitations would
give rise to massive particles that carry a scalar charge. The mass 
of the particles would be identified by locating the poles of the 
Feynman propagator in momentum space \cite{peskin}. In the absence of 
radiative corrections, and in flat spacetime, these particles would
have a mass that would stay constant, and this mass would directly
correspond to the mass parameter of the field's Lagrangian. The
physics of these elementary excitations might, however, be
substantially different in the curved spacetime of an expanding
universe. For example, Redmount \cite{redmount} has shown that in the
context of a real scalar field in de Sitter spacetime, the particle
energies are not constant --- they oscillate at late and early times
--- and do not correspond to the Lagrangian's mass parameter. Although 
Redmount's particles do not carry a scalar charge, his results clearly
suggest that the quantum physics of a scalar charge in an expanding
universe, even in the absence of radiative corrections, might be
considerably richer than the classical description provided in this  
paper. 

The second ingredient involved in a fundamental theory of radiating
scalar charges would be another scalar field $\Phi_{\rm radiation}$ 
coupled to the first to allow the scalar particles to radiate. This
new field would be real and massless, and it would generate the
radiative corrections that have so far been missing in our
description. These would modify the energies of the scalar charges
with respect to the (already complicated) free-field behavior. While 
this fully quantum description of a radiating scalar charge would
undoubtedly be richer and more interesting than the classical
treatment provided in this paper, we can hope that our classical  
considerations will not lead us too far astray --- they should indeed
provide us with a useful approximation. In this spirit we shall pursue
our speculations regarding the fate of scalar charges in our own
inflationary universe.  

For concreteness we assume that the time scale associated with the
inflationary epoch is $t_c \equiv |C| \sim 10^{-34}\ \mbox{s}$. This
produces a distance scale $r_c \equiv c t_c \sim 3 \times 10^{-24}\
\mbox{cm}$, a density scale $\rho_c \equiv 3/(8\pi G {t_c}^2) \sim 2
\times 10^{74}\ \mbox{g}/\mbox{cm}^3$, and a mass scale $m_c \equiv
4\pi \rho_c {r_c}^3/3 \sim 2 \times 10^{4}\ \mbox{g}$. For the
purpose of this discussion we reintroduce the speed of light
$c$ and the gravitational constant $G$, which were both previously set  
equal to one.     

We assume that a number $N$ of scalar charges, all of the same mass
$m_0$, are created prior to the onset of inflation, and that during
inflation, their mass decays according to Eq.~(\ref{6.18}). An
upper bound on $N$ is obtained by noting that at the onset of
inflation, $N m_0$ cannot exceed $m_c$, the total mass contained in
(what will become) the observable universe. Thus, $N < m_c/m_0$.  

Whether or not the scalar charges will survive the inflationary
epoch depends on the relation between $m_0$ and its duration 
$\Delta t$: if the particles are created heavy, and if inflation
doesn't persist for too long, then every particle present initially
will still be present after inflation. With our objective to produce
an upper bound for $N$, we assume that the scalar charges do survive  
the inflationary epoch, and we use Eq.~(\ref{6.18}) to derive a lower  
bound on the initial mass $m_0$. After inserting the appropriate
factors of $c$ and normalizing the (unknown) scalar charge $q$ to the
electron's charge $e$, we obtain 
\begin{equation}
m_0 > \frac{e^2}{c^3 t_c} \biggl( \frac{q}{e} \biggr)^2 
\biggl( \frac{\Delta t}{t_c} \biggr). 
\label{6.24}
\end{equation}
For an inflationary epoch that persists for approximately 60 e-folding  
times, we have that $m_0 > 5 \times 10^{-15} (q/e)^2\ \mbox{g}$. 
If $q/e$ is of order unity, this mass is larger by $6$ orders of
magnitude than the energy currently available at particle
accelerators.  

Combining this lower bound on $m_0$ with our previous result for $N$, 
we find that $N < 4 \times 10^{18} (e/q)^2$, so that the total number 
of scalar charges present today cannot exceed $10^{7} (e/q)^2$ per
galaxy. So we see that unless $e/q$ is huge, $m_0$ is naturally
extremely large, $N$ is naturally extremely small, and the prospect of
observing a scalar charge today is extremely limited.   

\begin{acknowledgments}

Our result for the retarded Green's function of a scalar field in de
Sitter spacetime, and the realization that scalar charges in de Sitter
spacetime must have a variable mass, were arrived at independently by
Amos Ori \cite{Ori}. We thank him for discussions, as well as Karel
Kucha\v{r}, Lee Lindblom, and Kip Thorne. L.M.B.\ wishes to thank the
Technion Institute of Theoretical Physics for hospitality. 

E.P.\ was supported by the Natural Sciences and Engineering Research
Council (Canada). A.I.H.\ was supported by a SURF fellowship. 
L.M.B.\ was supported by NSF grants PHY-0099568 (Caltech) and 
PHY-9734871 (University of Utah). 
\end{acknowledgments}

\begin{appendix}

\section{Alternative derivation of the scalar field Green's function for
de Sitter spacetime} \label{appa}

In this Appendix we present an alternative derivation for the retarded
Green's function of a scalar field in de Sitter spacetime.  

First, we observe that because of the four-dimensional homogeneity of
de Sitter spacetime, the Green's function can only be a function of
the invariant distance between two points. Next, we adopt coordinates
that make it easy for us to use this observation. Specifically, we
write the de Sitter metric in the form 
\begin{equation}
\,ds^{2}=-\,d\tau^{2}+\kappa^{-2}\sinh^{2}(\kappa\tau)\left(\,dr^{2}+
\sinh^{2}r \,d\Omega^{2} \right),
\label{a.1}
\end{equation}
where $\kappa=|C|^{-1}$ and $d\Omega^{2}$ is the standard metric on  
the unit two-sphere. In the coordinates of Eq.~(\ref{a.1}), the
spatial sections of de Sitter spacetime are open hyperboloids of
constant (negative) curvature. This form of the metric can be obtained
from the embedding relations 
\begin{eqnarray}
x_0&=&\kappa^{-1}\sinh(\kappa\tau)\cosh r\nonumber\\
x_1&=&\kappa^{-1}\cosh(\kappa\tau)\nonumber\\
x_2&=&\kappa^{-1}\sinh(\kappa\tau)\sinh r\cos\theta\label{co-trans}\\
x_3&=&\kappa^{-1}\sinh(\kappa\tau)\sinh r\sin\theta\cos\phi\nonumber\\
x_4&=&\kappa^{-1}\sinh(\kappa\tau)\sinh r\sin\theta\sin\phi\nonumber \, ,
\end{eqnarray}
which describe de Sitter spacetime as the hypersurface $-x_0^2 + x_1^2 
+ x_2^2 + x_3^2 + x_4^2 = \kappa^{-2}$ in a five-dimensional flat 
spacetime. The coordinate ranges are $-\infty<\tau<\infty$, $0\le
r<\infty$, $0\le\theta\le\pi$, and $0\le\phi\le 2\pi$. Notice that
these coordinates do not cover the entire de Sitter manifold. They go  
bad on $\tau=0$ and $r=\infty$, which correspond to two null rays
originating from the origin. More importantly, $\tau=0$ is a
coordinate singularity: all the points (with finite coordinate values)
$(\tau=0,r,\theta ,\phi)$ are the same physical point on the manifold
\cite{Ori}. This can most easily seen by direct substitution in the
coordinate transformation of Eq.~(\ref{co-trans}).  

These coordinates are particularly convenient, because surfaces of
constant geodesic distance from the origin are also surfaces of
constant $\tau$. We put the source at the origin of the coordinates
without loss of generality (because of the homogeneity of de Sitter
spacetime). Next, we will show that the proper time along a geodesic
between the points $P_{0}$ at $(0,0,0,0)$ and $P_{1}$ at
$(\tau,r,\theta,\phi)$ is simply $\tau$ for all finite values of
$(r,\theta,\phi)$. This is most elegantly accomplished by a method
suggested to us by Ori \cite{Ori}. From the previous result that all the
points with $\tau=0$ (and with finite coordinate values) are the same
point, $P_{0}$ may be represented by the coordinates
$(0,r,\theta,\phi)$, from which a constant $(r,\theta,\phi)$ curve
(which is a geodesic) may be extended to the point $P_{1}$ at
$(\tau,r,\theta,\phi)$. Next, observe that the proper time
along geodesics of constant $(r,\theta,\phi)$ is simply the coordinate
time. It then follows that the geodesic distance between $P_{0}$ and
$P_{1}$ is $\tau$. The retarded Green's function sourced at $(0,0,0,0)$ is
thus a function only of the coordinate time for all evaluation points,
that is $G(x ,0)=G(\tau ,0)=G(\tau)$.

Next, using the previous result that $G=G(\tau)$, we use Green's
equation to obtain an ordinary differential equation for
$G(\tau)$. Specifically, we find that  
\begin{equation}
\sinh^{-3}(\kappa\tau)\partial_{\tau}\left[\sinh^{3}(\kappa\tau)
\partial_{\tau}G(\tau)\right]=4\pi\frac{\delta_4(x)}{\sqrt{-g}}\, .
\end{equation}
Inside the source's light cone, the Green's function is a linear
combination of two linearly-independent solutions for the
corresponding homogeneous equation. These two solutions are just  
\begin{equation}
G_{1}(\tau)=1 \text{ and }
G_{2}(\tau)=\frac{\cosh(\kappa\tau)}{\sinh^2(\kappa\tau)}
+\frac{1}{2}\ln\frac{\cosh(\kappa\tau)-1}{\cosh(\kappa\tau)+1}\, .
\label{a.4}
\end{equation}
Hadamard's theory \cite{H} requires that
$G(\tau)=c_1G_1(\tau)+c_2G_2(\tau)$ be an analytic function. 
However, $G_2$ is not analytic: approaching $\tau=0$ it behaves like 
$G_2(\tau)=1/(2\kappa^2\tau^2)+\ln(\tau)/2+O(1)$. Hence, $c_2=0$, and we
conclude that the Green's function is a constant inside the source's
light cone.

The coordinates (\ref{a.1}) are inconvenient for finding the value of that
constant, because of their bad behavior on the light cone. Because the 
Green's function is a scalar, our conclusion that it is a constant
inside the light cone is unchanged when we switch to other
coordinates. As was discussed in Sec. \ref{green}, it may be directly
observed from the form of Green's equation in conformal coordinates
that the support on the source's light cone is identical to the
corresponding flat-space support. Transforming to the static
coordinates of Eq.~(\ref{6.16}), 
\begin{equation}
G(x ,0)=\kappa \coth(\kappa r^{*}) \delta(t-r^{*})+K\theta(t-r^{*}),
\label{a.5}
\end{equation}
where $K$ is yet to be determined.

Substituting this into the homogeneous 
Green's equation in these coordinates,
\begin{equation}
\left[\cosh^{2}(\kappa r^{*})\left(-\partial^{2}_{t}
+\partial^{2}_{r^{*}}\right)+2\kappa\coth(\kappa r^{*})
\partial_{r^{*}}\right]G(t,r^{*})=0,
\label{a.6}
\end{equation}
fixes $K=\kappa^{2}$. With this identification, 
Eq.~(\ref{a.5}) is equivalent to the Green's function of 
Eqs.~(\ref{4.1}) and (\ref{4.10}) adapted to de Sitter spacetime. 
\end{appendix}

\end{document}